\documentclass{article}

\usepackage{PRIMEarxiv}

\usepackage[utf8]{inputenc} 
\usepackage[T1]{fontenc}    
\usepackage{hyperref}       
\usepackage{url}            
\usepackage{booktabs}       
\usepackage{amsfonts}       
\usepackage{nicefrac}       
\usepackage{microtype}      
\usepackage{lipsum}
\usepackage{fancyhdr}       
\usepackage{graphicx}       
\graphicspath{{media/}}     

\usepackage[numbers,sort&compress]{natbib}
\usepackage{cleveref}
\title{Tetra-penta-deca-hexagonal-graphene (TPDH-graphene) hydrogenation patterns: dynamics and electronic structure
}

\author{
  Caique Campos, Matheus Medina, Pedro Alves da Silva Autreto \\
  Center for Natural and Human Sciences (CCNH) \\
  Federal University of ABC (UFABC) \\
  Santo André - SP, 09210-170, Brazil.\\
  \texttt{pedro.autreto@ufabc.edu.br} \\
   \And
  Douglas Soares Galvao \\
  Physics Institute Gleb Wataghin (IFGW) \\
  State University of Campinas (UNICAMP) \\
  Campinas/SP, Brazil\\
  \texttt{galvao@ifi.unicamp.br} \\
}

\begin{document}

\maketitle

\begin{abstract}
The advent of graphene has renewed the interest in other 2D carbon-based materials. Bhattacharya and Jana have proposed a new carbon allotrope, composed of different polygonal carbon rings containing 4, 5, 6, and 10 atoms, named Tetra-Penta-Deca-Hexagonal-graphene (TPDH-graphene). This unusual topology created material with interesting mechanical, electronic, and optical properties and several potential applications, including UV protection. Like other 2D carbon structures, chemical functionalizations can be used to tune their TPDH-graphene properties. In this work, we investigated the hydrogenation dynamics of TPDH-graphene and its effects on its electronic structure, combining DFT and fully atomistic reactive molecular dynamics simulations. Our results show that H atoms are mainly incorporated on tetragonal ring sites (up to $80 \%$ at\% at $300$ K), leading to the appearance of well-delimited pentagonal carbon stripes. The electronic structure of the hydrogenated structures shows the formation of narrow bandgaps with the presence of Dirac cone-like structures, indicative of anisotropic transport properties.
\end{abstract}

\section{Introduction}

The versatility in chemical bonding (different hybridizations) of carbon atoms allows the existence of a wide variety of different structures (allotropes) \cite{Diederich2010}, such as fullerenes \cite{Kroto1985}, nanotubes \cite{Iijima1991}, and graphene \cite{Novoselov2004}. Graphene is a 2D allotrope of $sp^{2}$ carbon atoms tightly packed into a hexagonal honeycomb lattice. It presents high carrier mobility ($5000 cm^{2}/V.s$ )\cite{Novoselov2004, Schedin2007}, high thermal conductivity ($5000 W mK^{-1}$) \cite{Balandin2008}, and Young modulus value of $1$ TPa \cite{Lee2008},  one of the highest values ever measured. It has unveiled new and unique physics phenomena, including the quantum Hall effect \cite{Zhang2005}, the ambipolar electric field effect \cite{Novoselov2004}, and the massless charge carriers of Dirac fermions \cite{Novoselov2005}. These remarkable properties have made graphene the subject of a large number of theoretical and experimental studies in different areas, such as catalysis \cite{Hu2017}, electronics \cite{Schwierz2010}, spintronics \cite{Han2014}, twistronics \cite{PhysRevB.95.075420}, and gas sensors \cite{C3TA11774J}, to name just a few.

However, graphene is a null electronic gap material, even exhibiting extraordinary electronic properties, which limits its use in some applications \cite{Novoselov2004}. Chemical functionalizations, such as hydrogenation, are one viable mechanism for altering graphene-like structures' properties (including opening the gap \cite{PhysRevB.75.153401, Marinho2021m, Lee2021} or changing the Fermi level \cite{doi:10.1021/acs.jpcc.7b01520}). Structural and electronic changes are introduced when the chemical species form covalent bonds. The partial hydrogenation of graphene introduces unsaturated $sp^{3}$ carbon atoms that can be used to attach additional functional groups. 

\begin{figure*}[h!]
    \centering
    \includegraphics[scale=1.1]{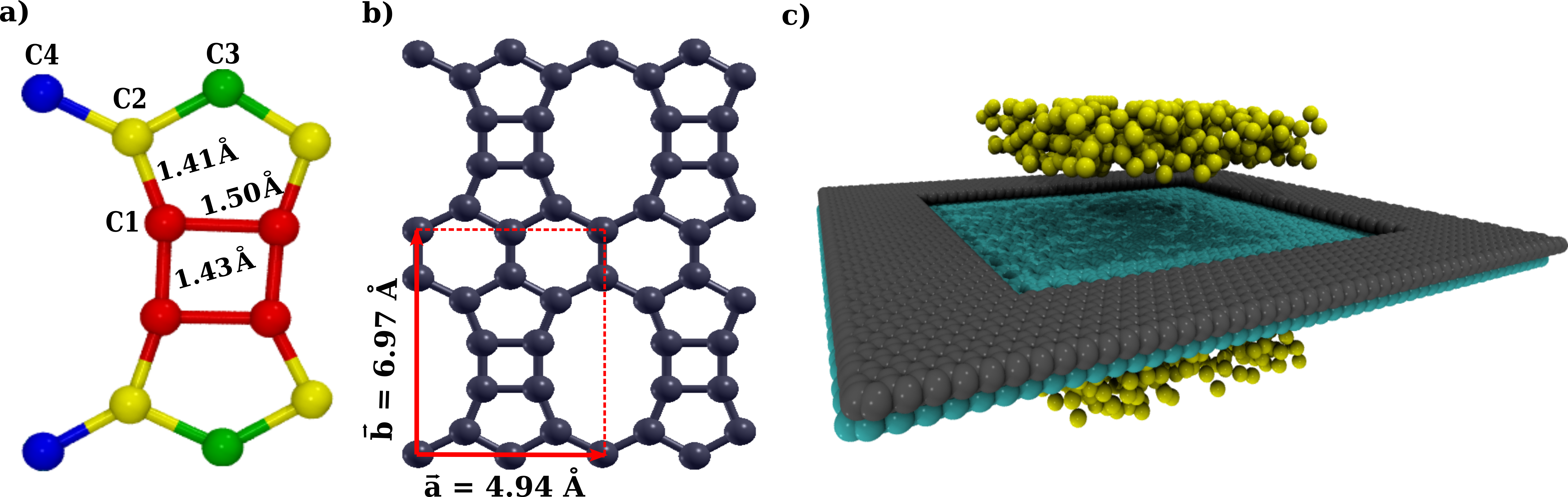}
    \caption{(a) Schematics of the unit cell of tetra-penta-deca-hexagonal-graphene (TPDH) and the corresponding carbon-carbon bond-length values. The different colors indicate non-equivalent carbon atoms. (b) A $2\times2$ supercell illustrating the TPDHG rings and the pores of the structure. The corresponding Unit cell vector values are indicated in the highlighted red rectangle. (c) The structural setup simulation used in the simulations.  A TPDH membrane (indicated in blue) is deposited on a graphene frame (gray), and the TPDHG/graphene structure is immersed in a hydrogen atmosphere (yellow). See text for discussions.}
    \label{fgr:TPDH-structure}
\end{figure*}

Despite these limitations, the advent of graphene created a revolution in materials science and renewed the interest in  2D carbon allotropes. Among these structures, it is worth mentioning graphynes and biphenylene carbon networks \cite{Baughman1987}. 

Graphynes are the generic name for families of 2D carbon porous structures containing hexagon rings connected by acetylenic groups and with $sp$ and $sp^2$ hybridized carbon atoms in the same lattice \cite{Baughman1987}. Graphdyines refer to the structural families where two acetylenic groups connect the hexagons \cite{Haley1997}. They can exhibit metallic and semiconducting behaviors \cite{PhysRevB.58.11009} and have been exploited in different technological applications \cite{Peng2014}.

Biphenylene carbon networks (including biphenylene carbon and graphenylenes) are families of porous structures composed of mixed carbon rings (pentagons, hexagons, heptagons, octagons, etc.) \cite{Baughman1987, Luo2021, Brunetto}. Similarly to graphynes, they can be metallic, or semiconductors and have potential applications in catalysis \cite{Luo2021}, gas sensors \cite{Hosseini2020}, batteries \cite{Yu2013}, and energy storage applications \cite{Hussain2017}. Recently, new synthetic routes for graphynes \cite{NatSynt,valentin} and biphenylene carbon networks \cite{sciencebpc} have been reported increasing the interest in these materials. Bhattacharya and Jana \cite{Bhattacharya2019} have proposed a new structure composed of two pentagons and a tetragonal ring called tetra-penta-octogonal graphene (TPO-graphene). It is metallic with a Dirac cone at $3.7$ eV above the Fermi level. More recently, they proposed another structure belonging to the tetra-pentagonal graphene family composed of $sp^2$ carbon rings with 4, 5, 12, and 6 atoms (Fig. \ref{fgr:TPDH-structure}) named tetra-penta-deca-hexagonal graphene (TPDH-graphene). It possesses thermal and dynamical stability and exhibits elastic anisotropy with Young's modulus value larger than that of graphene in a specific direction. Depending on the morphology, TPDH-graphene nanoribbons can exhibit metallic, or semiconductor behavior \cite{Bhattacharya2021}.



In this work, we have investigated the effects of hydrogenation on the structural and electronic properties of TPDH-graphene (TPDH-gr). The hydrogenation of TPDH-gr sheets was investigated through reactive molecular dynamics simulations. Structural optimization, energy, and electronic properties were further analyzed using ab initio (DFT) calculations.

In spite of graphene's extraordinary properties, it is a null gap material, which Chemical functionalization is one viable mechanism to introduce specific modifications into graphene-like structures. Structural and electronic changes are introduced when the chemical species being introduced form a covalent bond. For example, graphite oxides can form oxygen groups in graphene sheets dispersed in water and organic solvents \cite{Song2012}. Stankovich et al. prepared graphite oxides functionalized with isocyanates that were later exfoliated into graphene oxides dispersed in an aprotic polar solvent \cite{STANKOVICH20063342} in a stable manner. Partial hydrogenation of graphene sheets introduces unsaturated carbon atoms $sp^{3}$ that neighbor unpaired with electrons that can be used to attach additional functional groups. Chemical functionalization also allows one to change the electronic properties of the structure by opening a bandgap \cite{PhysRevB.75.153401, Marinho2021m, Lee2021} or changing the Fermi level \cite{doi:10.1021/acs.jpcc.7b01520}.

\section{Computational Methods}

First-principles calculations were carried out within the Density Functional Theory (DFT) framework as implemented in Quantum Espresso code \cite{Giannozzi_2009}. Electron-ion interactions were dealt with Projected Augmented wave (PAW) and Ultra-soft pseudopotentials for C and H atoms, respectively. They were obtained from the Standard Solid State Pseudopotentials library (SSSP) \cite{Prandini2018, Lejaeghere2016}. Exchange and correlation potential were used within the Generalized Gradient Approximation (GGA) with the parameterization of Perdew, Burke, and Ernzerhof (GGA-PBE functional) \cite{PhysRevLett.78.1396}. Valence electrons were treated with a set of plane waves basis set with a kinetic energy cutoff of 680 eV. The diagonalization of the density matrix was performed with the Davidson iterative method with matrix overlap using the self-consistency threshold of $10^{-6}$ eV. In the ionic relaxation calculations, the convergence thresholds were set to $10^{-3}$ eV and $10^{-2}$ eV/\AA $\:$ for energy and forces, respectively. Brillouin zone (BZ) sampling was performed using a $12\times12\times1$ ($16\times16\times1$) k-point grid for SCF (NSCF) calculations following the scheme proposed by Monkhosrt and Pack \cite{PhysRevB.13.5188}. For electronic structure calculations, the k-points were chosen along the following path in the BZ: $\Gamma(0,0,0)$ - $M(0.5, 0.5, 0)$ - $X(0.5, 0, 0)$ - $\Gamma(0, 0, 0)$ - $Y(0, 0.5, 0)$ - $M(0.5, 0.5, 0)$ - $\Gamma(0, 0, 0)$. 

We have also carried out fully atomistic molecular dynamics (MD) simulations using the large-scale atomic/molecular massively parallel simulator (LAMMPS) code\cite{plimpton1995fast}. Atomic interactions were treated with the reactive force field (ReaxFF) \cite{van2001reaxff}, with C-C interaction parameters developed by Chenoweth \textit{et al.}\cite{chenoweth2008reaxff}. All MD simulations were carried out in the canonical (\textit{NVT}) ensemble, with a time step of $0.25$ fs, and using a Nosé-Hoover thermostat \cite{hoover1985canonical}. The hydrogenation simulations were carried out considering a TPDH-gr membrane deposited on a graphene frame, as shown in Fig. \ref{fgr:TPDH-structure}.c. The TPDH-graphene membrane is a 24x15 supercell, in which only its central part (16x11) is exposed to the hydrogen atmosphere, resulting in a total number of 2112 available adsorption/reaction sites. The hydrogen atmosphere was composed of 500 atoms in a volume of 60 000 \AA$^3$ on each side of the membrane, constrained to the exposed region of the membrane. This methodology has been successfully applied to other systems, such as Me-graphane\cite{Marinho2021m} and graphone\cite{woellner_autreto_galvao_2016}.


\begin{figure*}[]
    \centering
    \includegraphics[scale=1.1]{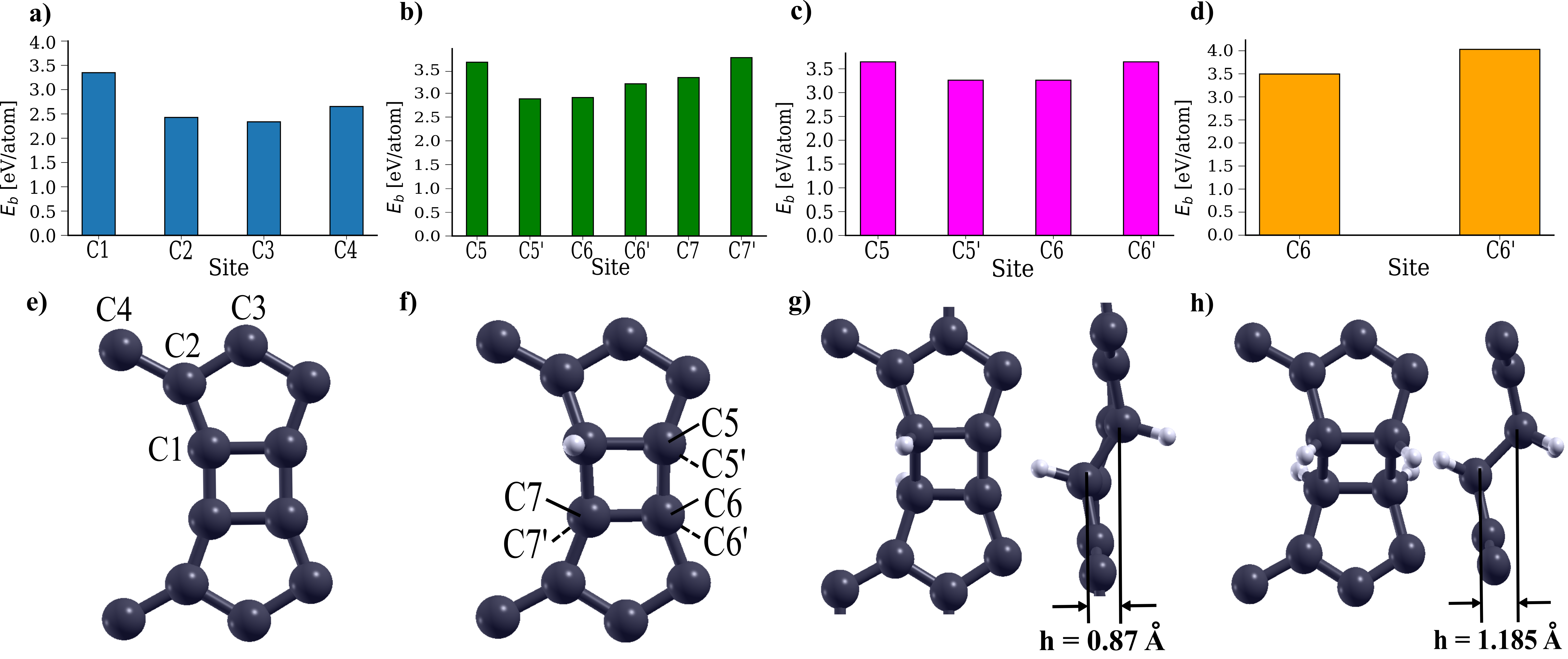}
    \caption{Adsorption energies for TPDH-gr in a) the non-equivalent sites, b) with an H atom adsorbed in the C1 site, c) two H atoms adsorbed in the C1 and C7' sites, and d) tree H atoms adsorbed in the C1, C7' and C5 sites. e) Non-equivalent sites in TPDH-gr. f) remaining sites in the tetragonal ring with the C1 site occupied. The top sites are indicated by the solid line, while the bottom sides are indicated by the dashed ones and prime labels. g) Top and side views of TPDH-gr with C1 and C7' sites occupied. The side view also shows the buckling height. h) Top and side views of TPDH-gr with a fully hydrogenated tetragonal ring.}
    \label{fgr:binding-energies}
\end{figure*}

\section{Results and Discussion}

\subsection{Ab initio Binding Energy and Hydrogenation Dynamics}
TPDH-gr has a $Pmmm$ (space group $\# 47$) symmetry; the $12$ carbon atoms in its unit cell are arranged in an orthorhombic lattice. The obtained optimized lattice parameters were: $a = 4.94$ \AA, $b = 6.97$ \AA $\:$ with $\gamma = 90^{o}$. There are three different bond lengths ($1.41$, $1.50$, and $1.44$ \AA $\:$.) involving the C atoms, as shown in Fig. \ref{fgr:TPDH-structure}.a. Except for the C atoms bonded along the $\vec{a}$ direction in the tetragonal ring, the bond lengths are close to those $sp^2$ in graphene ($1.41$ \AA)\cite{C2RA22664B}. These results agree well with those reported by Batthacharya and Jana \cite{Bhattacharya2021}.

The most favorable sites for H adsorption/reaction were investigated by evaluating the \textit{binding energy} per adsorbed atom, calculated as the energy difference between the hydrogenated structure and its parts:

\[ E_b = - \left[ \frac{E_{TPDH+nH} - (E_{TPDH} + nE_{H})}{n} \right] \]

\noindent where $E_{TPDH+nH}$ is the energy of TPDH-gr with n adsorbed H atoms, $E_{TPDH}$ is the energy of a TPDH-gr unit cell, and $E_{H}$ the energy of an isolated H atom. The negative sign means that high energies indicate more favorable sites for adsorption than others in the same structure. First, an H atom is adsorbed at each of the non-equivalent sites (Fig. \ref{fgr:TPDH-structure}.a). The site corresponding to the highest energy is taken as the most favorable. Then, a second H is adsorbed at each of the remaining sites, and the most favorable one is evaluated according to $E_b$. This process is repeated until the tetragonal ring on the TPDH-gr is fully hydrogenated. We present the binding energies and obtained structures in Fig. \ref{fgr:binding-energies}.

The adsorption of the first H atom is more favorable on the $C1$ site,  as seen in Fig. \ref{fgr:binding-energies}.a with $E_b$ of $3.35$ eV/atom. After C1-Cx adsorption (with x = 2, 5, and 7), the bond length values increased to $1.51$, $1.55$, and $1.53$ \AA $\:$, respectively, indicating a transition to $sp^{3}$=like-bond in the C1 atom.  It is worth mentioning that for sites located in the tetragonal ring, the top and bottom configurations (Fig. \ref{fgr:TPDH-structure}.d) were considered. Adsorption of a single H atom at each of these sites resulted in roughly the same results for $E_b$, as can be seen in Table 1S in Supplementary Material. 

The adsorption of the second H atom (resulting in 16\% hydrogen coverage) is more favorable at the $C7'$ site (Fig. \ref{fgr:binding-energies}.c), with an $E_b$ of $+3.75$ eV/atom. The resulting lattice distortions in the direction perpendicular to the structure plane lead to a significant buckling of $h = 0.87$ \AA $\:$,  as seen in Fig. \ref{fgr:binding-energies}.f. The distortions of the structure and the fact that two neighboring C atoms adsorb the pair of H atoms (but on opposite sides of the sheet) are in accordance with the results reported by Boukhvalov and Katsnelson for the hydrogenation of graphene sheets \cite{Boukhvalov_2009}.

Interestingly, the adsorption of a third H atom gives the same $E_b$ for both C5 and C6' sites, as seen in Fig. \ref{fgr:binding-energies}.e. In this case, the configuration in which the C1, C7' and C5 sites are occupied was imposed, which will be justified later. The resulting structure presents an overall increase in the Cx-C bond lengths (with x = 1, 7, and 5). The vertical distance separating the C1 and C7 atoms is 1.02 \AA $\:$ versus 0.84 \AA $\:$ for the corresponding value between the C5 and C6 atoms. The adsorption of a fourth H atom (33\% hydrogen coverage) is more favorable at the C6' site with $E_b = 4.0$ eV/ \AA $\:$ and buckling of $h = 1.185$ \AA (Fig. \ref{fgr:binding-energies}.g, h). It is clear that choosing the C5 or C6' sites in the adsorption of the third H atom leads basically to the same configuration (C1-C7'-C5-C6'). Therefore, choosing C5 or C6' for the adsorption of the third H atom is equivalent. 

These calculations reveal a pattern for the hydrogenation of the tetragonal ring, which consists of two lines of H atoms on opposite sides of the basal plane sheet, leading to the formation of well-delimited pentagonal ring strips along the direction of the lattice vector $a$. DFT calculations confirm that this configuration is indeed more favorable. Molecular Dynamics simulations, discussed below, produced similar results,

\begin{figure*}
    \centering
    \includegraphics[width=16cm]{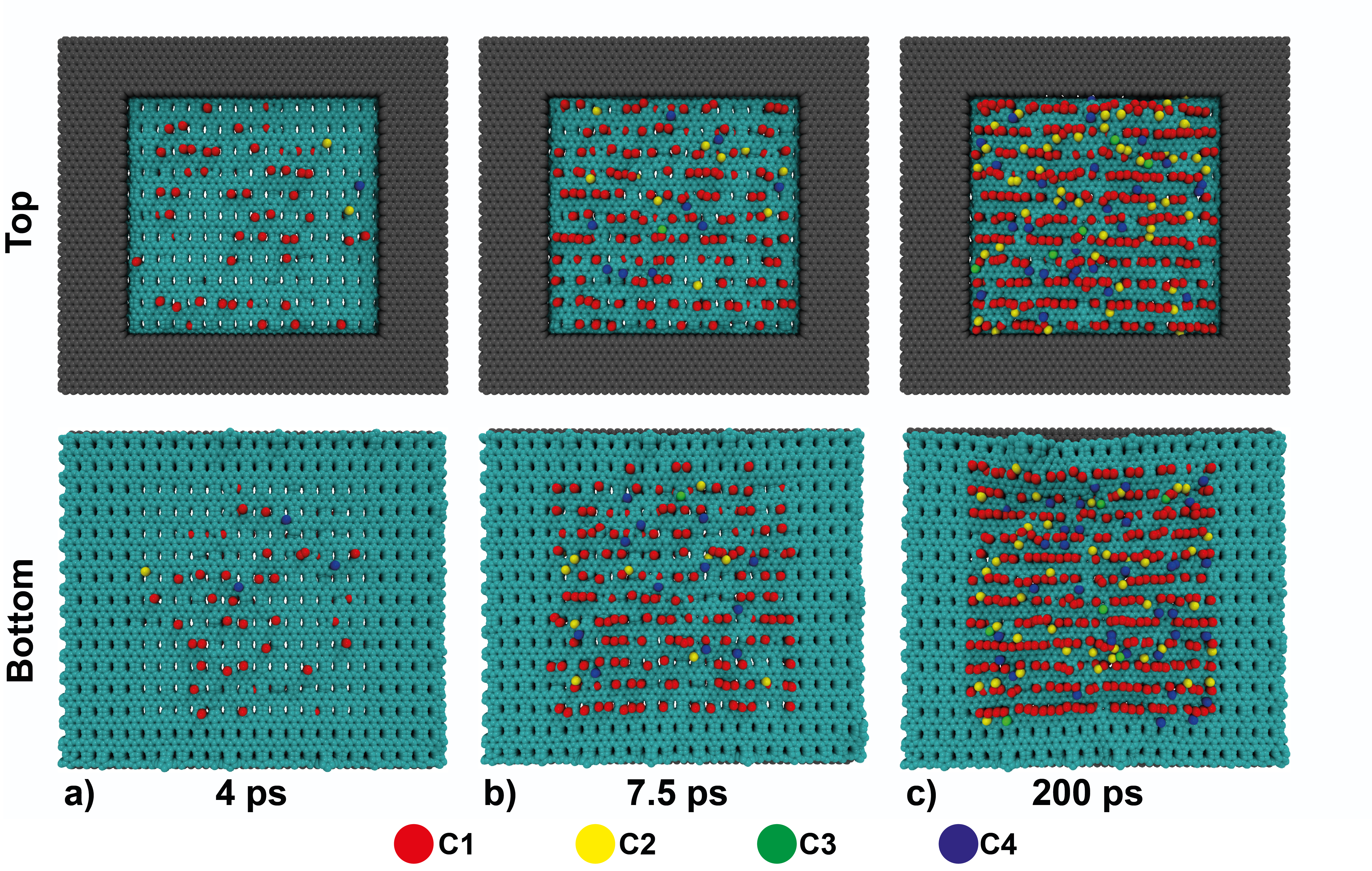}
    \caption{Representative MD snapshots at different simulation times: a) 4 ps, b) 7.5 ps, and c) 200 ps of the hydrogenation of the TPDH-gr membrane. Results from simulations at 300k.}
    \label{fgr:TPDH-Hydro}
\end{figure*}

Reactive molecular dynamics simulations were carried out to study the dynamics and temperature effects on hydrogen adsorption of bigger TPDH-graphene membranes (Fig. \ref{fgr:TPDH-structure}), which would be cost-prohibitive with DFT methods. Representative MD snapshots of both sides of the TPDH-graphene membrane during the hydrogenation process (at 300K) are presented in Fig. \ref{fgr:TPDH-Hydro} (a) - (c).

The H atoms are predominantly incorporated throughout the MD simulations on the $C_1$ sites. Analyzing the hydrogenation process, from Fig. \ref{fgr:TPDH-Hydro} (a) to (c), we can see that the hydrogen-adsorbed $C_1$ sites act as seeds to the hydrogenation of their $C_1$ neighbors, forming lines through the structure surface, which is an expected result, based on the DFT binding energy ordering values. 

\begin{figure}
\centering
  \includegraphics[width=0.5\textwidth]{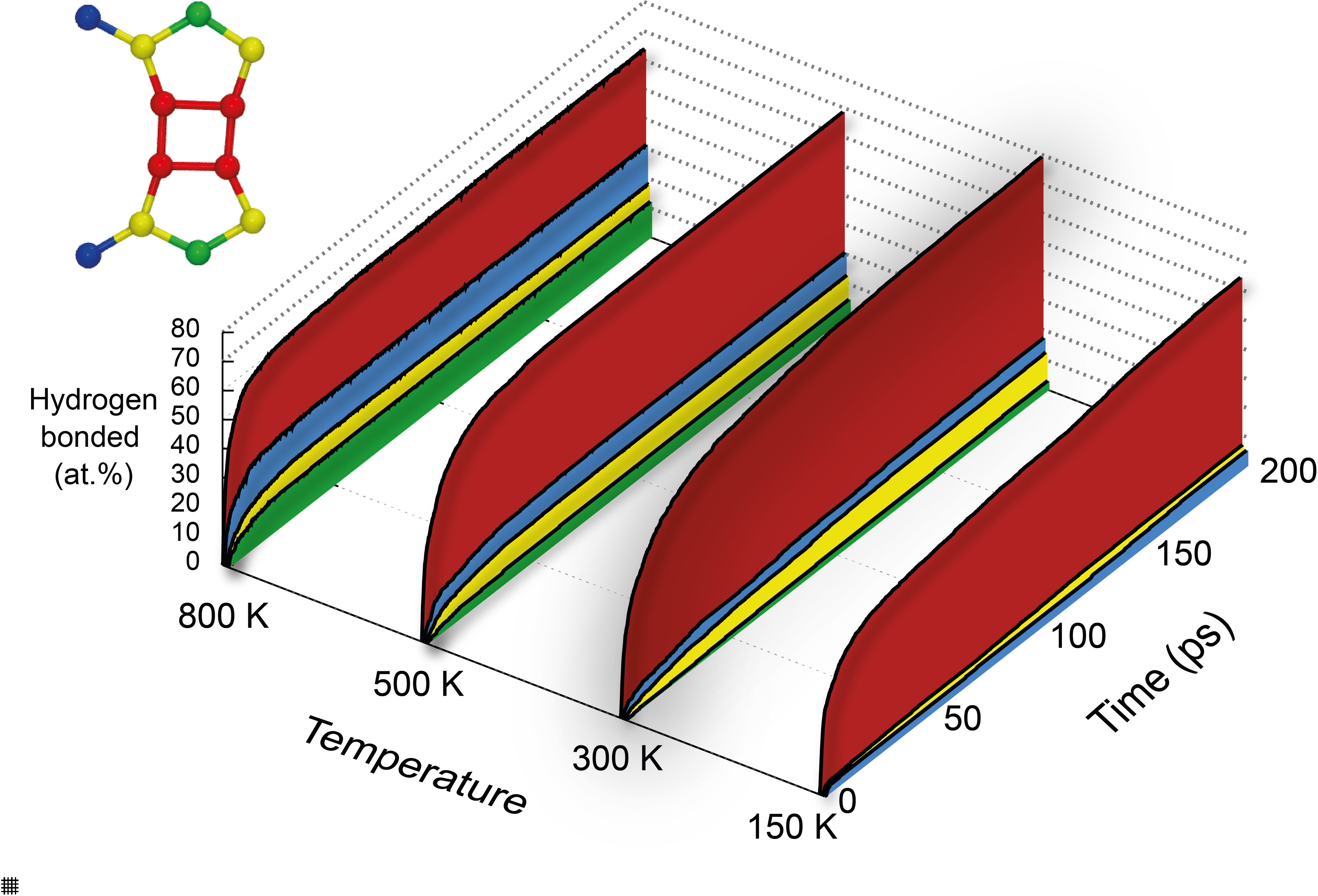}
  \caption{The number of adsorbed/bonded hydrogen atoms at each site of the TPDH-gr unit cell as a function of the simulation time (at \%) for 150, 300, 500, and 800 K. The color of the curves indicates the corresponding sites in the unit cell (left, upper).}
  \label{fgr:TPDH-ALLad}
\end{figure}

In Fig. \ref{fgr:TPDH-ALLad}, we present the number of adsorbed/bonded hydrogen atoms at each site of the TPDH-gr unit cell, as a function of the simulation time, for the different temperature values considered here. The hydrogenation occurs mainly at the $C_1$ sites for all temperatures. High rates of H incorporation indicate high reactivity for hydrogenation. At low temperatures (150K), the $C_2$ and $C_4$ sites have approximately the same low adsorption rates, while the $C_3$ sites exhibit insignificant or no hydrogen incorporations. Increasing the temperature, $C_4$, $C_2$, and $C_3$ sites become more reactive, while above 300K, the $C_1$ site has a slight decrease in reactivity.

\subsection{Electronic Structure}

\begin{figure}[t]
\centering
  \includegraphics[scale=1.8]{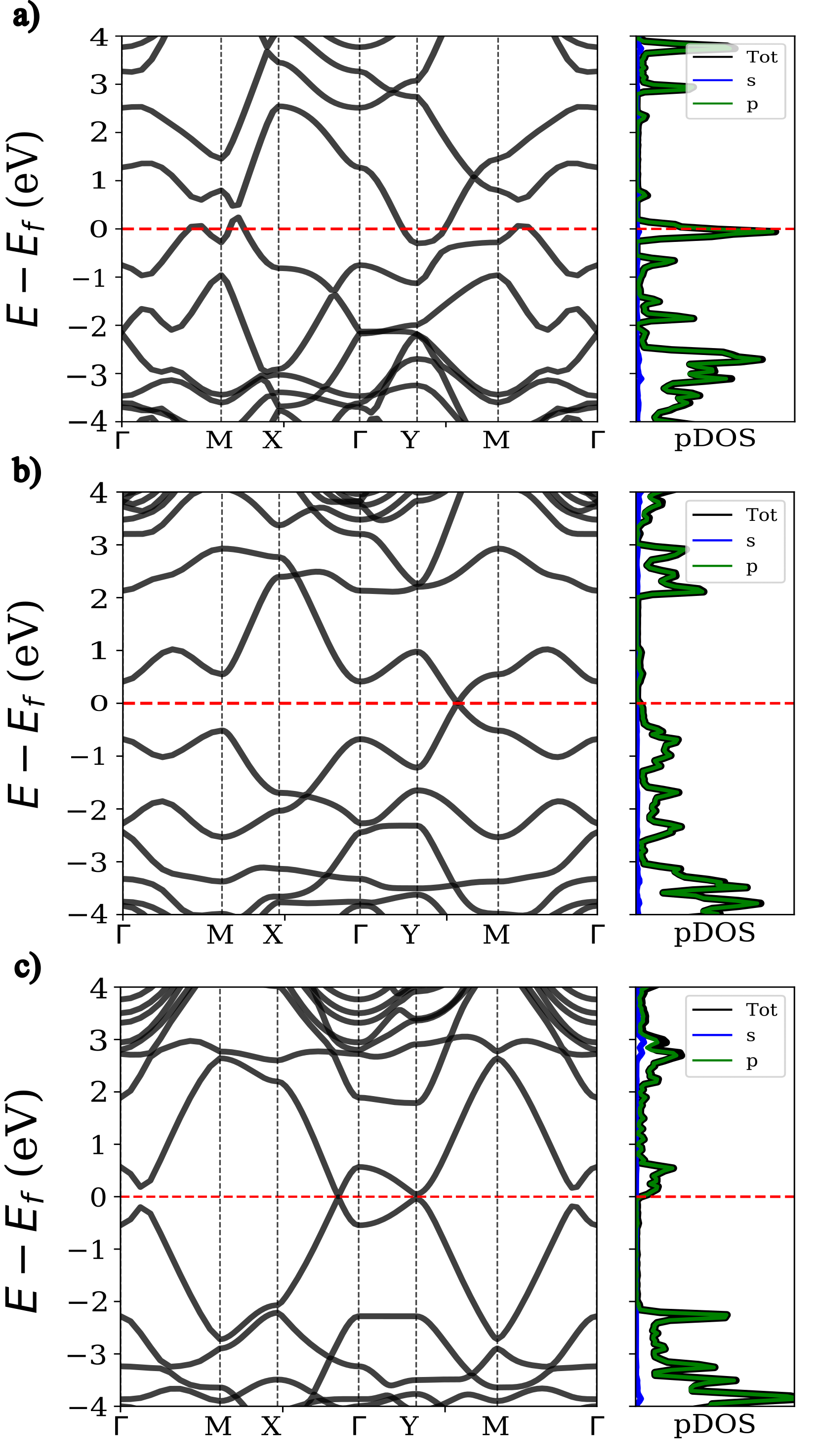}
  \caption{Electronic band structures and the corresponding projected density of states (pDOS) for a) non-hydrogenated TPDH-gr, b) TPDH-gr with the tetragonal ring partially hydrogenated (C1 and C7' sites occupied), and c) tetragonal ring fully hydrogenated. The total density of states is shown in black, while the blue and green curves represent the projected DOS into orbitals s and p, respectively.}
  \label{fgr:bands}
\end{figure}

In Fig. \ref{fgr:bands}.a), we present pristine (non-Hydrogenated) TPDH-gr electronic band structure and the corresponding projected density of states (pDOS) (obtained from DFT-GGA-PBE calculations). We can see that pristine TPDG-gr exhibits a semimetallic behavior. The highest (lowest) valence (conduction) band is partially filled. These results are consistent with previous works published in the literature \cite{Bhattacharya2021}. 

The effects of H adsorption in the tetragonal ring were investigated for the cases with a pair adsorbed in neighboring atoms, in opposite sites of the sheet, and with all four sites of the ring occupied (Fig. \ref{fgr:binding-energies}.g,h respectively).

The adsorption of two hydrogen atoms in the C1 and C7' sites results in the opening of the direct gaps by approximately $1$ eV at k-points $M$ and $\Gamma$, as shown in Fig. \ref{fgr:bands}.b. Surprisingly, the valence and conduction bands overlap at the Fermi level, giving rise to a Dirac cone-like, between the k-points $Y$ and $M$. Near this point, the electronic dispersion is unusually linear, and charge carriers behave like massless fermions, obeying the Dirac relativistic equation. It is expected that unusual transport properties arise from this pattern in the band structure, as predicted and experimentally observed for graphene \cite{Novoselov2005}. The electronic band structure and the corresponding pDOS of the TPDH with full hydrogenation of the tetragonal ring are shown in Fig. \ref{fgr:bands}.c. We can see the appearance of narrow gaps ($0.5$ eV ) between k-points $\Gamma$ and $M$, and a very narrow direct gap at point $Y$. The Dirac cone-like is shifted near the $\Gamma$ points with respect to the half-hydrogenated structure.

\section{Conclusions}

This work investigated the effects of hydrogenation on the structural and electronic properties of tetra-penta-deca-hexagonal-graphene (TPDH-gr) sheets. Molecular dynamics (MD) simulations revealed that H atoms are mainly incorporated in the tetragonal ring ($C_1$ sites) with up to $80\%$ adsorption at $300$ K (Fig. \ref{fgr:TPDH-ALLad}). The number of H atoms incorporated on C2 and C4 sites varies according to the temperature. Hydrogenation produces a pattern where H lines are formed on both sides of the sheet (Figs. \ref{fgr:TPDH-structure}.h and \ref{fgr:TPDH-Hydro}.c) generating well delimited pentagonal ring strips along $\vec{a}$ direction. DFT calculations further corroborate that the complete hydrogenation of the tetragonal ring is energetically favorable. 

Electronic structure calculations for the partially hydrogenated structure show the formation of gaps and the emergence of a Dirac cone-like between the points $\Gamma$ and $M$. For the fully hydrogenated ring, narrow band gaps followed by wide gaps are identified, and the Dirac cone-like is translated near the $\Gamma$ point. This electronic profile strongly indicates anisotropic transport properties, although these remain to be further explored in future works.


\section*{Conflicts of interest}

There are no conflicts to declare.

\section*{Acknowledgements}


The authors thank PRH.49 for funding and CCM-UFABC for the computational resources provided and CNPq (\#310045/2019-3)





\bibliography{main} 

\end{document}